# Journal Quality Factors from ChatGPT: More meaningful than Impact Factors?

Mike Thelwall
Information School, University of Sheffield, UK. https://orcid.org/0000-0001-6065-205X
m.a.thelwall@sheffield.ac.uk
Kayvan Kousha
Wolverhampton Business School, University of Wolverhampton, UK. https://orcid.org/0000-0003-4827-971X

**Purpose**: Journal Impact Factors and other citation-based indicators are widely used and abused to help select journals to publish in or to estimate the value of a published article. Nevertheless, citation rates primarily reflect scholarly impact rather than other quality dimensions, including societal impact, originality, and rigour. In response to this deficit, Journal Quality Factors (JQFs) are defined and evaluated. These are average quality score estimates given to a journal's articles by ChatGPT.
**Design/methodology/approach**: JQFs were compared with Polish, Norwegian and Finnish journal ranks and with journal citation rates for 1,300 journals with 130,000 articles from 2021 in large monodisciplinary journals in the 25 out of 27 Scopus broad fields of research for which it was possible. Outliers were also examined.
**Findings**: JQFs correlated positively and mostly strongly (median correlation: 0.641) with journal ranks in 24 out of the 25 broad fields examined, indicating a nearly science-wide ability for ChatGPT to estimate journal quality. Journal citation rates had similarly high correlations with national journal ranks, however, so JQFs are not a universally better indicator. An examination of journals with JQFs not matching their journal ranks suggested that abstract styles may affect the result, such as whether the societal contexts of research are mentioned.
**Research limitations:** Different journal rankings may have given different findings because there is no agreed meaning for journal quality.
**Practical implications**: The results suggest that JQFs are plausible as journal quality indicators in all fields and may be useful for the (few) research and evaluation contexts where journal quality is an acceptable proxy for article quality, and especially for fields like mathematics for which citations are not strong indicators of quality.
**Originality/value**: This is the first attempt to estimate academic journal value with a Large Language Model.

## 1  Introduction

Academics and research managers seem to be often concerned with Journal Impact Factors (JIFs) or other journal citation rate formulae to assess the value of published articles or to help select a journal to publish in. For example, a tenure committee might check the JIF of a previously unknown journal that a candidate has published in (McKiernan et al., 2019) or a researcher might consult journal rankings when deciding where to submit their work. Although there are arguments (Hecht et al., 1998; Seglen, 1997) and initiatives against the overuse of journal indicators as a proxy for article quality, such as the Declaration on Research Assessment (DORA, sfdora.org) to promote more responsible research assessment practice (Curry et al., 2022), they may still sometimes play a useful role, such as when the assessors lack the time or expertise to critically evaluate an article or when the time necessary for this outweighs the informational value of an accurate decision. Whilst some argue that article citation rates should be used in this situation, they are not available for new articles and journal citation rates can sometimes be better indicators of article quality since they suggest the degree of rigour in the journal reviewing process (Waltman & Traag, 2020) and correlate positively with article quality in all fields (Thelwall et al., 2023).

Following from the above point, one of the criticisms of JIFs is that citations (sometimes) reflect academic impact and do not directly reflect the other important dimensions of journal article quality, including societal impact, rigour, and originality (Langfeldt et al., 2020; Aksnes et al., 2019; Wouters, 2014). This has been an unavoidable problem until recently, when the emergence of large language models has made automatic peer review (Bharti et al., 2024; Fiorillo & Mehta, 2024; cf. Suleiman et al., 2024), reporting standards assessment (Roberts et al., 2023) and quality evaluation (de Winter, 2024; Thelwall & Yaghi, 2024) for academic research more possible, adding to the many uses of Artificial Intelligence (AI) within broader review systems (Kousha & Thelwall, 2024). There is now strong preliminary evidence that ChatGPT can make reasonably accurate estimates of the quality of journal articles (Thelwall & Yaghi, 2024) by following the UK Research Excellence Framework (REF) expert assessor guidelines (REF2021, 2019). It is thus logical to assess whether it is possible to design indicators for journals that are based on automatic assessments of all dimensions of quality to replace the current scholarly impact-based journal citation rate indicators.

This article introduces the Journal Quality Factor (JQF), which is defined to be the average quality score of the qualifying articles published in a given year. In theory, the quality score could be derived from any source, including human experts and Artificial Intelligence (AI), but ChatGPT will be used in the current paper because it has been shown to correlate positively with expert scores all fields (Thelwall, 2024ab; Thelwall & Yaghi, 2024; Thelwall et al., 2024). Despite the availability of various journal ranking lists, there is no gold standard for journal quality and so JQF is assessed instead by comparing it to available national journal rankings and a known phenomenon, journal citation rates. The research questions are as follows.
- RQ1: In which fields do ChatGPT-based JQF values correlate positively with national journal ranks and with journal citation rates?
- RQ2: Why do ChatGPT-based JQF values differ from national journal ranks for individual journals (in the sense of being scatterplot outliers)?

## 2 Methods

The research design was to obtain up to 100 large monodisciplinary journals in all Scopus broad fields from 2021, then to calculate journal citation rates, national journal ranking scores, and ChatGPT JQFs for the journals, correlating them and examining anomalies for insights. Small journals with fewer than 100 articles in a year were excluded to avoid unreliable JQFs. The year 2021 was selected to give three years of citations for the citation rate formula, which seems to be sufficient for reliable results (Wang, 2013). Multidisciplinary journals (defined as those classified within at least two different broad Scopus fields) were also excluded to allow field differences in the results to be identified. Scopus was chosen in preference to other citation databases, such as the Web of Science (WoS), Dimensions, and OpenAlex, because its classification scheme (27 broad fields) is at the appropriate level of granularity. Correlation was used as the primary analysis mechanism because this allows the new indicator, JQF, to be benchmarked against a known indicator type and journal rankings.

## 2.1 Journal ranks

Although there are some international journal ranks for individual fields and most countries do not rank journals, a few have such lists and make them public in a usable format. In terms of less useful lists, Excellence in Research Australia used tiered rankings in 2010 but discontinued them, Qualis in Brazil is partly based on JIFs, Italian national journal ratings are available only for scientific areas, Pakistan's lists for science and social science are only available in PDF form, which is error-prone to scan, and South Africa has an accredited journal list without ratings. In contrast, Finland's expert rankings of the Julkaisufoorumi (julkaisufoorumi.fi; Pölönen et al., 2021; Saarela & Kärkkäinen, 2020), the expert rankings of the Norwegian Register for Scientific Journals (Pölönen et al., 2021), and Poland's bibliometric and expert hybrid rankings of the Lista czasopism punktowanych (gov.pl, 2024; Pölönen et al., 2021; for an earlier version, see: Kulczycki & Rozkosz, 2017) are publicly available lists of academic journals with ratings or points and so these were used, when they matched Scopus journal names, for comparison by correlation. All rankings were downloaded on the 6$^{th}$ of October 2024.

- Polish journal ranks: These were obtained from https://www.gov.pl/web/nauka/komunikat-ministra-nauki-z-dnia-05-stycznia-2024-r-w-sprawie-wykazu-czasopism-naukowych-i-recenzowanych-materialow-z-konferencji-miedzynarodowych, with points (20, 40, 70, 100, 140, 200=highest) on 10 October 2024, and were produced by expert teams from the academic community for the Minister of Science.
- Finnish journal ranks: These were obtained from https://jfp.csc.fi/jufoportaali, with ranks (0, 1, 2, 3=highest) and were produced by over 300 researchers in 23 expert panels organised by the Publication Forum.
- Norwegian journal ranks: These were obtained from https://kanalregister.hkdir.no/publiseringskanaler/AlltidFerskListe with ranks (0, 1, 2 = highest) and were produced by the National Board of Scholarly Publishing with the help of scientific panels.

Journal ranking lists were merged by name(s) and ISSN(s) and then matched with Scopus journals.

## 2.2 Dataset and citation counts

The sample of journal articles was identified as follows. First, all Scopus journal articles were downloaded from Scopus using its API in January-February 2024. Second, all journals in this dataset were identified by name. Third, journals with articles in more than one Scopus broad field (or classified as Multidisciplinary) were excluded.

To obtain the input for ChatGPT, the titles and abstracts of the Scopus articles were used. The abstracts were first processed to remove copyright statements and other publisher standard information (Webometric Analyst at github.com/MikeThelwall/Webometric_Analyst; Citations menu, Remove copyright statements…). Although a previous study also removed structured abstract terms (Thelwall & Yaghi, 2024), they were retained here because the prior study got better results in one case without excluding them. The titles and abstracts were then combined into a single line, with "\nAbstract\n" in the middle, as the ChatGPT input. Previous research has shown that

ChatGPT's quality estimates from titles and abstracts correlate more highly with human quality scores than when using full text input (Thelwall, 2024b; Thelwall & Yaghi, 2024). After this, articles with an abstract length above 785 characters were selected, discarding the 25% of articles with shorter abstracts. This is because some journals publish short form articles with short abstracts and an examination of the data also showed many shorter abstracts from non-article submissions, such as corrections. Thus, this minimum length requirement ensures a more consistent set of articles to examine.

Next, journals with fewer than 100 qualifying articles were excluded. This left only one journal in Decision Sciences, which was therefore omitted, because more are needed to calculate a correlation coefficient. The remaining fields had at least six. For fields with more than 100 journals, a random sample of 100 was taken to be a sufficiently large set for comparisons. Finally, 100 articles were sampled at random (using a random number generator) from each of the remaining journals. This dataset included Scopus citation counts. The final sample size was 1,300 journals containing 130,000 articles in total, although not all journals had ranks from each country (Table 1).

Table 1. Number of journals with ranks in each country for all non-Multidisciplinary Scopus fields.

| Field | Finland | Norway | Poland | Any | Journals |
|---|---|---|---|---|---|
| Agricultural and Biological Sciences | 81 | 76 | 94 | 94 | 100 |
| Arts and Humanities | 4 | 4 | 5 | 5 | 5 |
| Biochemistry, Genetics & Molecular Biology | 93 | 90 | 97 | 97 | 100 |
| Business, Management and Accounting | 14 | 17 | 17 | 17 | 17 |
| Chemical Engineering | 12 | 14 | 16 | 16 | 19 |
| Chemistry | 59 | 56 | 61 | 61 | 64 |
| Computer Science | 76 | 75 | 83 | 84 | 86 |
| Decision Sciences [not included] | 1 | 1 | 1 | 1 | 1 |
| Earth and Planetary Sciences | 81 | 80 | 89 | 89 | 100 |
| Economics, Econometrics and Finance | 16 | 16 | 17 | 17 | 18 |
| Energy | 23 | 22 | 26 | 26 | 31 |
| Engineering | 67 | 65 | 90 | 91 | 100 |
| Environmental Science | 60 | 62 | 65 | 66 | 70 |
| Immunology and Microbiology | 13 | 11 | 13 | 13 | 15 |
| Materials Science | 38 | 37 | 46 | 47 | 50 |
| Mathematics | 31 | 30 | 34 | 34 | 34 |
| Medicine | 82 | 84 | 93 | 93 | 100 |
| Neuroscience | 35 | 34 | 35 | 35 | 35 |
| Nursing | 21 | 22 | 24 | 24 | 25 |
| Pharmacology, Toxicology & Pharmaceutics | 47 | 47 | 58 | 58 | 61 |
| Physics and Astronomy | 86 | 85 | 90 | 90 | 91 |
| Psychology | 32 | 32 | 32 | 32 | 32 |
| Social Sciences | 75 | 74 | 83 | 83 | 86 |
| Veterinary | 22 | 23 | 25 | 25 | 25 |
| Dentistry | 25 | 26 | 30 | 30 | 30 |
| Health Professions | 5 | 5 | 6 | 6 | 6 |
| All | 1099 | 1088 | 1230 | 1234 | 1301 |

## 2.3 ChatGPT quality scores

The selected article titles and abstracts were submitted to ChatGPT 4o-mini via its API. At the time, the two main versions were 4o and 4o-mini. They seem to give similar results for this type of task (Thelwall, 2024b), so 4o-mini was chosen as the cheaper option.

With each article submitted to ChatGPT, system instructions were also sent to define the academic journal article quality scoring task. These instructions were the same as previously used (Thelwall & Yaghi, 2024), which seems to give better results than a more multidimensional approach (de Winter, 2024) and are the Research Excellence Framework (REF) quality scoring guidelines for expert assessors, reformulated into a standard ChatGPT instruction format. There are four guidelines, corresponding to health and life sciences (A), physical sciences and engineering (B), social sciences (C), and the arts and humanities (D). For each field, the matching ChatGPT system guideline was used (by Scopus code A: 11, 13, 24, 27, 28, 29, 30, 32, 34, 35, 36; B: 15, 16, 17, 18, 19, 21, 22, 23, 25, 26, 31; C: 14, 20, 33; D: 12).

Each ChatGPT output included a REF quality score (1*, 2*, 3*, 4* or a number between these values) as part of an overall report on the submitted article. These REF scores were extracted by a program designed for this (Webometric Analyst's AI menu, ChatGPT: extract

REF scores option), calling on human adjudication for reports from which it could not find the score. In cases where ChatGPT reported separate scores for originality, significance, and rigour, but not an overall score, these three were averaged.

### 2.4   JQF and journal citation rate (JCR)

JQFs were calculated for each journal as the arithmetic mean of the ChatGPT scores for the 100 selected articles. Although more precise scores can be obtained for individual articles by averaging multiple scoring attempts from ChatGPT (Thelwall, 2024b), this was not necessary because the focus is on journal averages.

Journal citation rates were calculated for each journal as the geometric mean of the citation counts of the 100 selected articles (i.e., the same set of articles as for the ChatGPT scores). The geometric mean is superior to the arithmetic mean for highly skewed sets of numbers, such as article citation counts (Thelwall & Fairclough, 2015) and is close to the median (Vogel, 2022). Some journal impact indicators used field (and year) normalised citation counts (e.g., Leydesdorff & Opthof, 2010; Moed, 2016) but this was not necessary since each set of journals is monodisciplinary.

### 2.5   Analyses

JQF were compared with journal citation rates for each field using Pearson correlations. This measures the extent to which the relationship between the two is linear. In contrast, Spearman correlations were calculated between JQFs and the Finnish, Polish, and Norwegian journal ranks. Although the Finnish (0, 1, 2, 3) and Norwegian (0, 1, 2) systems have simple levels, the Polish is numerical and skewed (20, 40, 70, 100, 140, 200) so Spearman correlations were used for all. Journals that were not listed in any of the systems were treated as missing data. Since no single country is of greatest importance here, the median of the three correlations was used to represent the overall journal rank correlation.

Journals that do not fit the trend are important as they may give insights into factors that may lead ChatGPT to give an inappropriate score. These were identified for qualitative investigation by choosing the five fields with the lowest correlations, excluding small fields (<10 journals). These may include the most anomalous journals. Anomalies were found by drawing scatter plots of ChatGPT scores against journal points from Poland (which had the widest range of levels) and visually identifying anomalies in the sense of journals breaking the main trend. These were then investigated by reading a sample of ChatGPT reports to try to discover plausible reasons for the anomalies. Although metric-based formulae for selecting anomalies are available, the visual approach was preferred here to identify journals that were also close to the extremes of the distribution, since they would be easier to manually investigate.

## 3   Results

The quantitative results are reported first, followed by the qualitative analysis of outliers, in the fields suggested by the quantitative analysis.

### 3.1   JQF and journal citation rate correlation

JQFs have positive median correlations with the Polish, Finnish and Norwegian journal ranks in 24 out of the 25 fields investigated, and the correlation is statistically significantly different from 0 in 23 (Figure 1). Recall that two fields were excluded: Multidisciplinary for spanning multiple fields, and Decision Sciences for having only one qualifying journal. The correlations are generally high, with 23 being at least 0.5.

Journal citation rates and JQFs tend to have similar correlations with the Polish, Finnish and Norwegian journal ranks within each field (Figure 1). The main exception is Mathematics, where JQFs correlate highly with national journal ranks (median correlation: 0.824) but JCRs have a correlation of 0. Although the differences are less extreme, there is a statistically significantly higher JQR correlation for Economics, Econometrics and Finance; Agricultural and Biological Sciences; and Biochemistry, Genetics and Molecular Biology. Conversely, there is a significantly lower JQR correlation only for Computer Science.

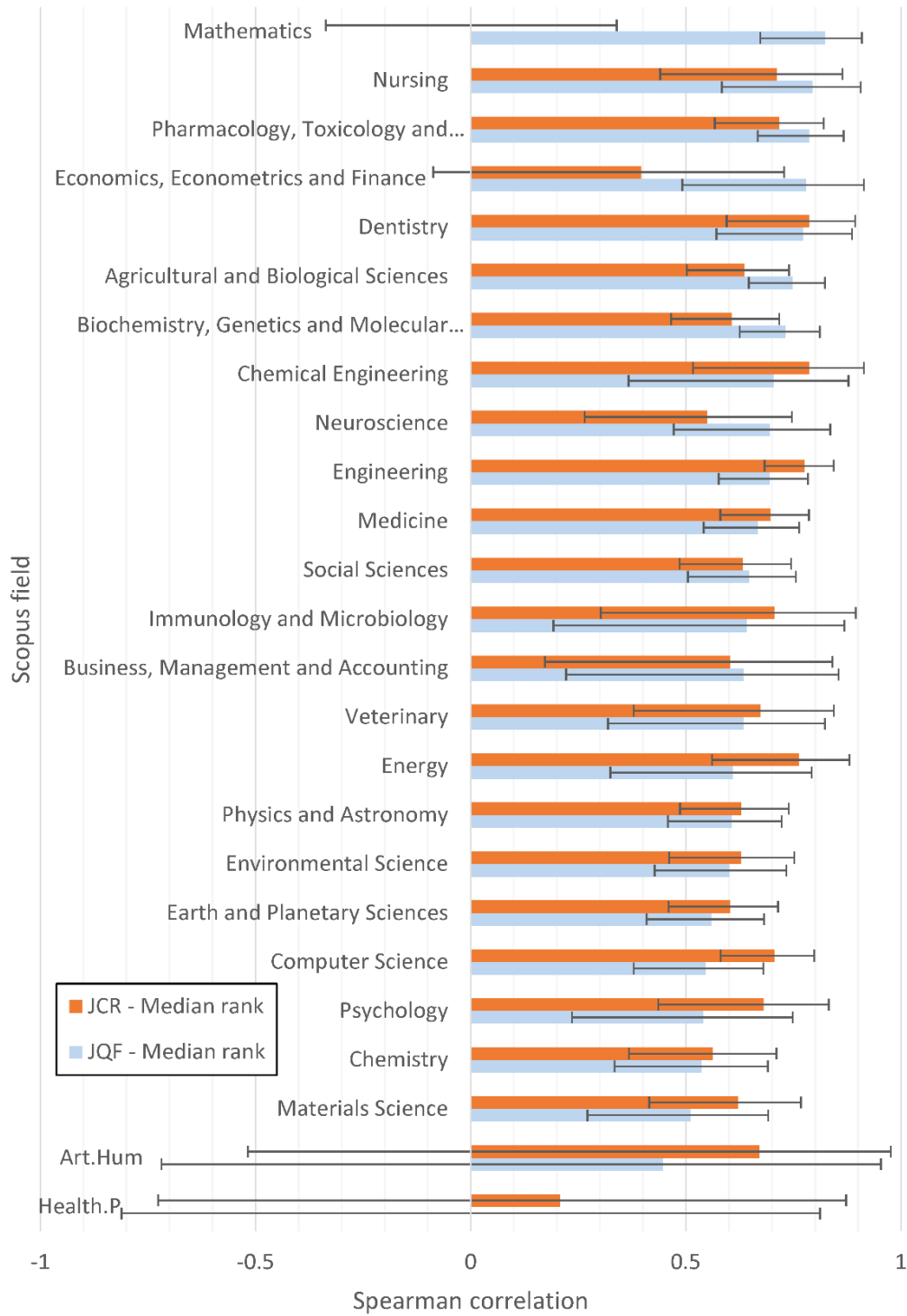

Figure 1. Median correlations between Polish/Finnish/Norwegian journal ranks and either JQFs or journal citation rates (JCRs) for all Scopus broad fields except Multidisciplinary and Decision Sciences. Each field is represented by up to 100 monodisciplinary journals with a random sample of 100 articles each. Error bars represent 95% confidence intervals.

The correlations between JQFs and the three different national rankings are broadly similar within each individual field. The main exceptions are fields represented by few journals, presumably due to larger error margins for these (Figure 2). The same is true for the correlations between JCRs and the three different national rankings (Figure 3).

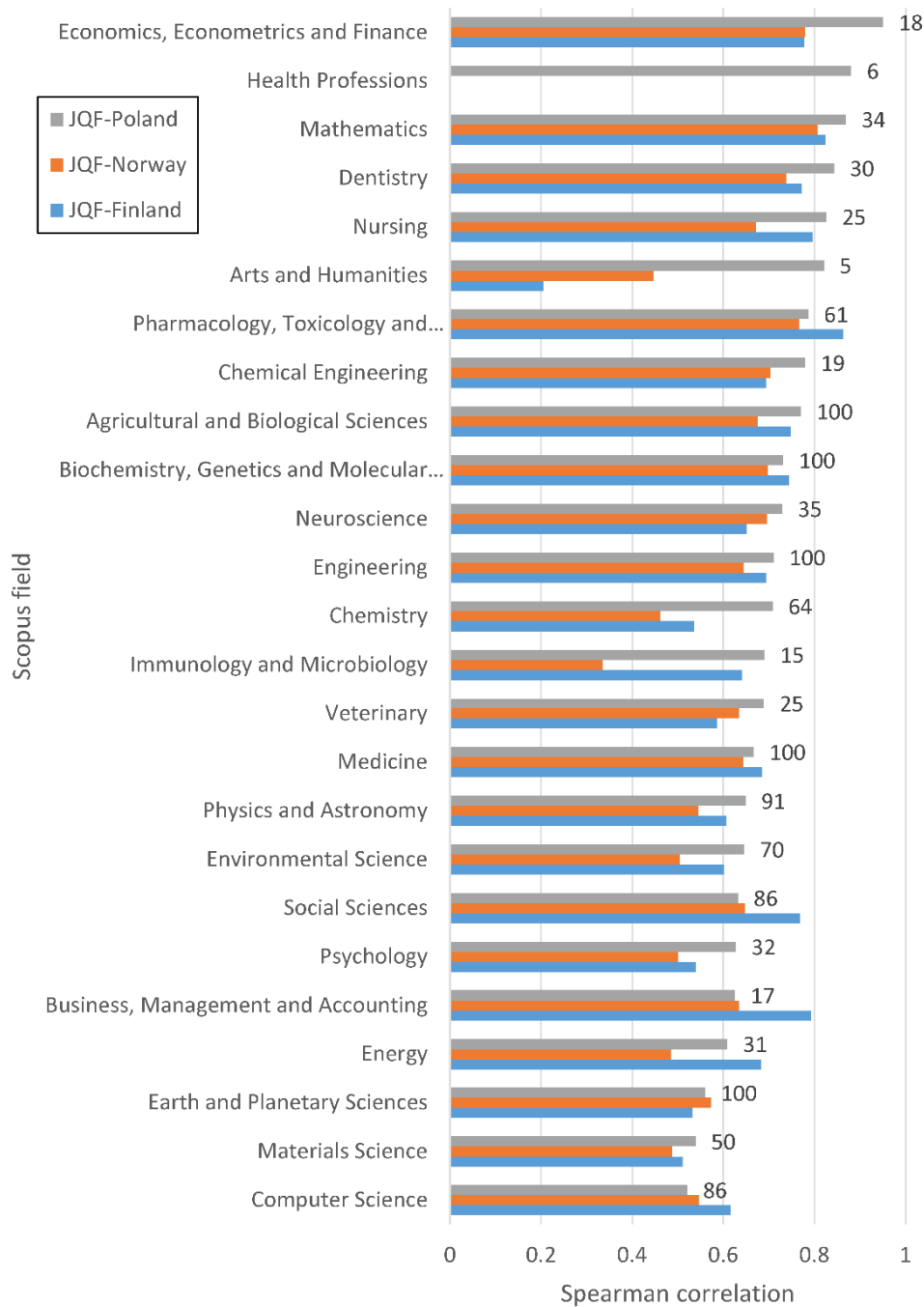

Figure 2. Correlations between Polish, Finnish and Norwegian journal ranks and JQFs for all Scopus broad fields except Multidisciplinary and Decision Sciences. Each field is represented by up to 100 monodisciplinary journals (as stated next to the bars) with a random sample of 100 articles each.

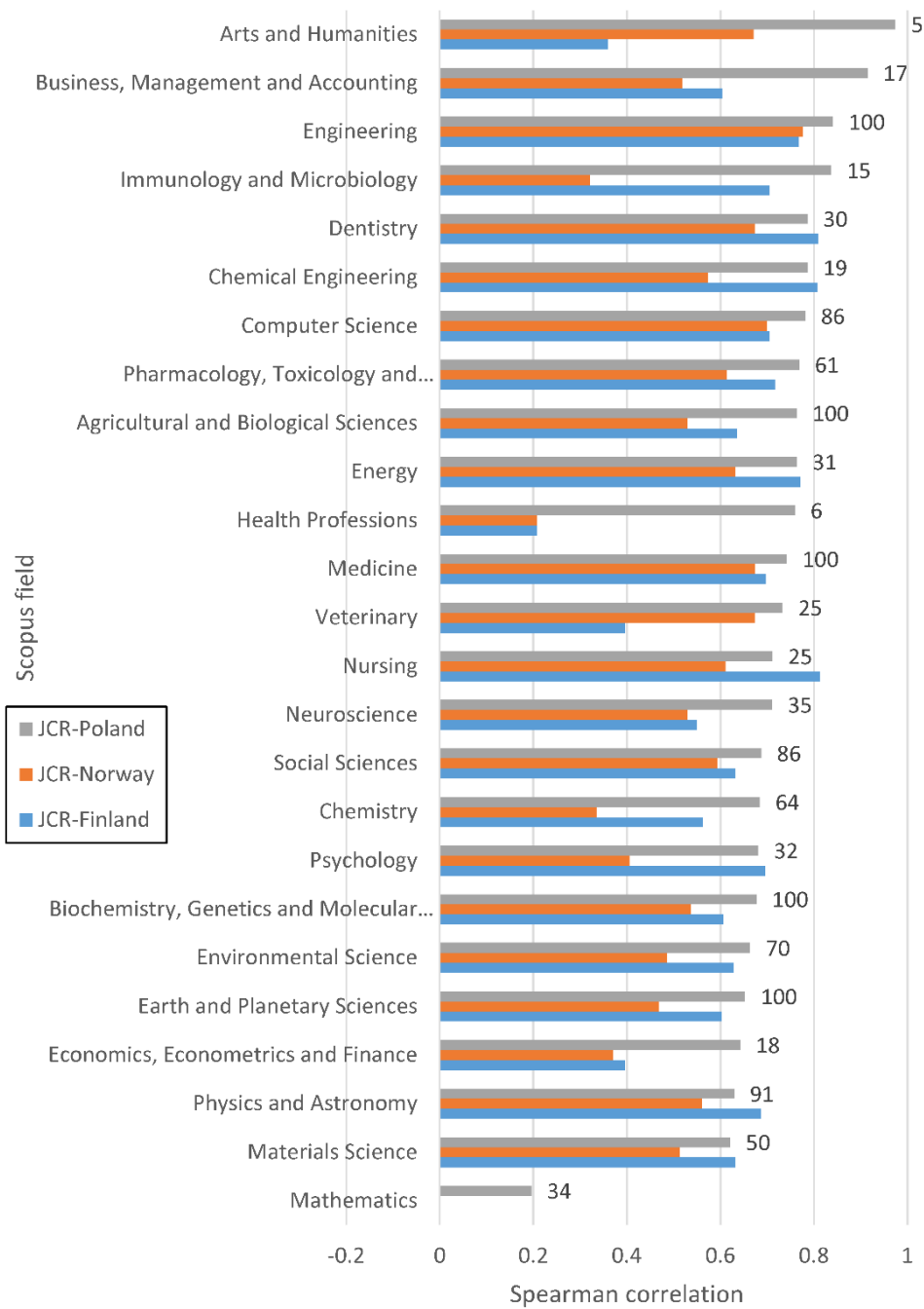

Figure 3. Correlations between Polish, Finnish and Norwegian journal ranks and journal citation rates (JCRs) for all Scopus broad fields except Multidisciplinary and Decision Sciences. Each field is represented by up to 100 monodisciplinary journals (as stated next to the bars) with a random sample of 100 articles each.

Overall, both JCRs and JQFs correlate most strongly with the Polish rankings and the Polish ranking is the only one for which JCRs correlate more highly than JQFs (Table 2). Speculatively, perhaps the Polish rankings for the journals covered take national factors into account less than the other two rankings.

Table 2. Median Spearman correlations between JQFs or JCRs and national citation ranks across all Scopus broad fields except Multidisciplinary and Decision Sciences.

|  | Finland | Norway | Poland |
|---|---|---|---|
| **JQF** | 0.682 | 0.635 | 0.708 |
| **JCR** | 0.632 | 0.537 | 0.733 |

All Scopus broad fields had a positive correlation between JQFs and journal citation rates, and it was statistically significant in 24 out of 25 examined (Figure 4). The correlations are generally very high. All except one is at least 0.5, and 15 out of 25 are at least 0.75. It is surprising that the correlations are high even for areas where citation counts are known to be weak or very weak indicators of individual article research quality, such as the Arts and Humanities (albeit for only 5 journals) and Social Sciences (Nederhof, 2006). The relatively high correlations can be partly explained as an aggregation effect of 100 articles per journal, however.

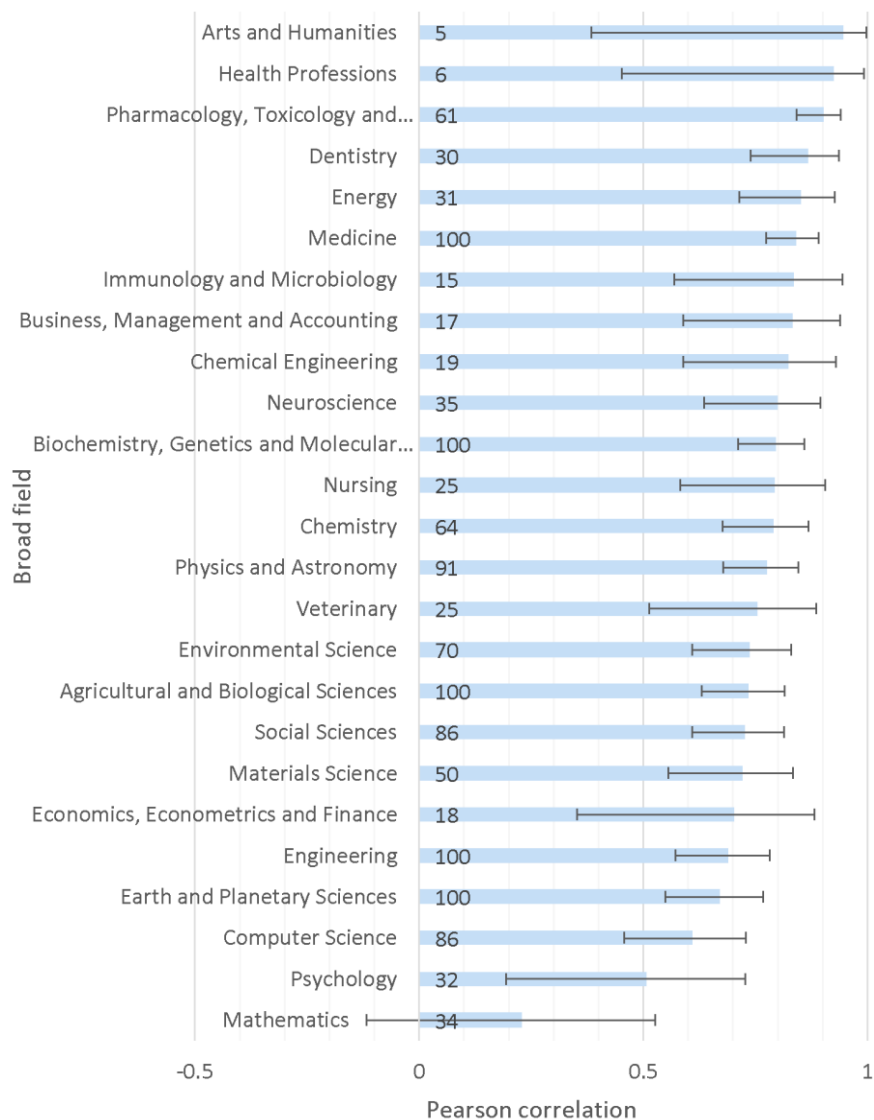

Figure 4. Correlations between JQFs and journal citation rates for all Scopus broad fields except Multidisciplinary and Decision Sciences. Each field is represented by up to 100

monodisciplinary journals with a random sample of 100 articles each. Numbers on bars record the number of journals for the field. Error bars represent 95% confidence intervals.

### *3.2 Trend-breaking journals in low correlation fields*

From scatter graphs of the five non-small fields with the lowest correlation (Materials Science, Chemistry, Psychology, Computer Science, and Earth and Planetary Science), nine journals were identified and investigated for having anomalously high or low ChatGPT average scores for their Polish journal rank. The journals with anomalously low ChatGPT scores are discussed first.

- *Journal of Natural Fibers* has a low JQF of 2.43 (7[th] lowest in Materials Science) but mostly high journal ranks (Finland: 2, Norway: 1, Poland: 140): This might be due to its relatively short abstracts (10[th] lowest in Materials Science) but it seems more likely to be related to the relatively applied and specific nature of its topics (e.g., "Dyeing of Mulberry Silk Using Binary Combination of Henna Leaves and Monkey Jack Bark", "A Study on Fatigue Behavior of Dual Core-Spun Yarns Containing Wool and Elastane Cores"). Its abstracts also seem to be less precise and more exploratory than those of *Matter* (discussed below).
- *Journal of Attention Disorders* has a low JQF of 2.59 (lowest in Psychology) but mostly high journal ranks (Finland: 2, Norway: 2, Poland: 140): This journal may have had lower ChatGPT scores due to a lack of explicit non-academic context setting in abstracts (e.g., "Objective: This study aimed to investigate differences in parent- and teacher-reported ADHD symptoms according to the child's country of origin.[…]", "Objective: The objective of this study was to compare psychiatric comorbidity and consumption-related variables in ADHD patients seeking treatment for cocaine, cannabis, or both.[…]","Objective: This study was the first attempt to explore the efficacy of a mindfulness protocol for children with attention-deficit hyperactivity disorder (ADHD) and oppositional defiant disorder (ODD), and their parents.[…]").
- *Journal of Autism and Developmental Disorders* has a low JQF of 2.71 (4[th] lowest in Psychology) but mostly high journal ranks (Finland: 2, Norway: 1, Poland: 140): As in the above case, social context information was rarely explicit in abstracts (e.g., "We conducted a cross-sectional study to explore whether clinical characteristics and autism diagnostic-traits severity are associated with caregiver-reported impairment of health-related quality of life (Pediatric Quality of Life Inventory Generic Core Scales, Fourth Edition) in 93 Chilean toddlers (age: 2–4 years) with autism spectrum disorder. […]", "We examined special education classifications among students aged 3–21 in North Carolina public schools, highlighting autism spectrum disorder (ASD) and intellectual disability (ID).[…]").
- *RISTI - Revista Iberica de Sistemas e Tecnologias de Informacao* has a low JQF of 2.17 (lowest in Computer Science) but a moderate journal rank (Finland: -, Norway: -, Poland: 70): Its low scores may be due to the indirect, imprecise and perhaps wordy nature of its abstracts (e.g., "The present still does not allow us to talk about the past, but it already allows us to reflect on some consequences of COVID-19 in society two years after its emergence. Internal communication was initially quite affected by the confinement that forced teleworking (office home) leading companies to a necessary adaptation with new means of communication in the management of internal relations.[…] Using this paper we try to present the reality imposed on companies that, starting with trying to avoid crisis situations, end up finding an opportunity for the development of internal communication through more updated and effective means with the use of digital tools, which we believe can enhance the consistency of the corporate brand.").

The journals with anomalously high ChatGPT scores are discussed next.

- *Matter* has a high JQF of 3.61 (highest in Materials Science) but medium or low journal ranks (Finland: 1, Norway: 1, Poland: 20): Its high ChatGPT scores may be for the highly technical nature of its abstracts (e.g., "In the field of skin-attachable electronics, debonding-on-demand (DoD) adhesives triggered by mild, efficient, and accessible stimuli can facilitate repeated usage with negligible damage to the skin. Here, a simple and versatile method has been developed to fabricate biocompatible bonding/debonding bistable adhesive polymers (BAPs) with skin temperature-triggered conformal adhesion and room temperature-triggered easy detaching.[…]").
- *JACS Au* has a high JQF of 3.48 (3rd highest in Chemistry) but medium or low journal ranks (Finland: 1, Norway: 1, Poland: 20): This journal seems to have concise technical abstracts with initial motivation sentences (e.g., "Mixing transition metal cations in nearly equiatomic proportions in layered oxide cathode materials is a new strategy for improving the performances of Na-ion batteries. The mixing of cations not only offers entropic stabilization of the crystal structure but also benefits the diffusion of Na ions with tuned diffusion activation energy barriers. In light of this strategy, a high-rate Na0.6(Ti0.2Mn0.2Co0.2Ni0.2Ru0.2)O2cathode was designed, synthesized, and investigated, combining graph-based deep learning calculations and complementary experimental characterizations. […]").
- *CCS Chemistry* has a high JQF of 3.61 (2nd highest in Chemistry) but medium journal ranks (Finland: 1, Norway: -, Poland: 40): Like *Matter*, this seems to have concise and technical abstracts (e.g., "The interaction between isolated transition-metal atoms and neighboring dopants in single-atom catalysts (SACs) plays a key role in adsorption strength tuning and catalytic performance engineering. Clarifying the local coordination structures of SACs is therefore of great importance and yet very challenging at the atomic level. Here, we employ a SAC with isolated Pt species anchored on nitrogen-doped carbon as a prototype and investigate the local coordination environment around Pt sites with the CO probe molecule by combined electrochemical infrared (IR) spectroscopy and density functional theory calculations. […]").
- *Proceedings of the ACM on Interactive, Mobile, Wearable and Ubiquitous Technologies* has a high JQF of 3.25 (highest in Computer Science) but high, medium, and low journal ranks (Finland: 2, Norway: 1, Poland: 20): Its abstracts are terse, motivating and precise (e.g., "Breathing biomarkers, such as breathing rate, fractional inspiratory time, and inhalation-exhalation ratio, are vital for monitoring the user's health and well-being. Accurate estimation of such biomarkers requires breathing phase detection, i.e., inhalation and exhalation. However, traditional breathing phase monitoring relies on uncomfortable equipment, e.g., chestbands. […] This paper assesses the potential of using smartphone acoustic sensors for passive unguided breathing phase monitoring in a natural environment. […]"). This journal may have been misclassified by Poland or not considered important for Polish research.
- *ACS Earth and Space Chemistry* has a high JQF of 3.16 (18th highest out of 100 in Earth and Planetary Sciences) but medium or low journal ranks (Finland: 1, Norway: 1, Poland: 20): This has impressive sounding abstracts and it not clear why it was low ranked by Poland (e.g., "Moderately volatile elements (MVEs) are variably depleted in planetary bodies, reflecting the imprints of nebular and planetary processes. [...] To quantitatively understand why Na, K, and Rb are depleted in planetary bodies, we have carried out vacuum evaporation experiments from basaltic melt at 1200 and 1400 °C to study their evaporation kinetics and isotopic fractionations. [...]").

Although this analysis suggests that the anomalies might be caused by "good" journals being disadvantaged by ChatGPT for having abstract styles that do not encourage a consideration of the wider value of the research, and "weak" journals benefitting from highly technical abstracts with societal motivations, it is also possible that the journal rankings are incorrect in the above cases and it is also possible that other factors that are the real underlying cause were not identified.

## 4   Discussion

The study is limited by the sample selection, including the monodisciplinary journal definition. More interdisciplinary journals might follow a different pattern, as might the data for different quality definitions in ChatGPT. The results might also vary by publication year, LLM, and ChatGPT version. The JCR correlations reported above are probably higher than correlations for JIFs since the JCR uses a more precise central tendency measure. The quality indicators used are also relatively coarse-grained, which may tend to reduce the correlations. Journals also change over time, such as due to editorial shifts, and the ranking schemes may have a temporal mismatch with the data. From the perspective of the reason for the JQF correlations, ChatGPT may have ingested the journal ranks and has somehow used them in its results. Finally, the evidence from the anomaly analysis is speculative and subjective.

The positive correlations between the new JQF and the existing JCR have multiple interpretations. It is possible that both are reasonable indicators of journal quality (i.e., the average quality of the articles in the journal) in most fields and the correlations reflect this. The mostly strong positive correlations between three national journal ranks and JQFs or JCRs are also consistent with this. Whilst the latter provides the strongest evidence in theory, it seems likely that JIFs or similar citation-based impact indicators are influential for the Finnish, Norwegian and Polish rankings. Thus, the results do not provide strong evidence that JQFs directly reflect the average quality of articles in journals.

The disciplinary comparisons give evidence that JQF is a distinctly superior quality indicator than JCRs in one and possibly four fields, with the reverse being true in one. It is perhaps surprising that the JQF correlations with national journal ranks were not always higher than the JCR correlations with journal ranks, given that ChatGPT scores can reflect all three quality dimensions, whereas JCRs only directly reflect scholarly impact. Although there is no direct evidence, it may be due to the JIF influencing national journal rankings, as mentioned above. Thus, this suggests that the JQF might be a better overall journal quality indicator than the JCR and therefore the JIF. Conversely, however, ChatGPT may have a limited ability to detect rigour, significance, and/or originality, especially because it is probably dependent on how authors describe these in their abstracts, if at all.

## 5   Conclusion

The results show that a journal quality indicator based on averaging ChatGPT article scores gives similar results to journal citation rates, in the sense that both tend to correlate positively and strongly with three national journal quality rankings (all of which may have been informed by journal citation rates). They also suggest that the JQF might be particularly useful for mathematics, where journal citation rates have little association with perceived research quality. In contrast, the analysis of anomalous journals suggested that some may be disadvantaged for ChatGPT by encouraging abstracts that ignore the societal value of studies. In the era of national research assessments and in the context of important research evaluations for jobs not necessarily being undertaken by field specialists, this seems unfortunate.

The use of journal-based indicators as a surrogate for article-level quality in any context is controversial because of a frequent overreliance on them and their possible overall detrimental impact on science. Nevertheless, as discussed in the introduction, there are some valid uses, and scientometricians and research evaluators should be aware of these to make an informed choice about whether it is appropriate to use a journal-based indicator in any particular case. This decision should consider the possible negative systemic effects caused by emphasising publication venues in a research evaluation (Rushforth & Hammarfelt, 2023; Wilsdon et al., 2015). In this context, the JQF, if implemented, may improve the overall informational value of journal-based indicators in some fields, and particularly mathematics. It seems unlikely that JQFs will replace JIFs because the latter are well embedded and the former lacks transparency and is somewhat ephemeral (relying on a particular technology in a fast-evolving area). Moreover, if too much value is given to JQFs then an unintended consequence (i.e., systemic effect) that might occur is gaming by journals by encouraging authors to generate ChatGPT-friendly abstracts that may even obscure the real value of a study by exaggerating claims and downplaying limitations.

Another important practical issue is that it seems reasonable to hypothesise that ChatGPT may be less accurate for, and perhaps biased against, work written by people that are not native English speakers. Thus, JQFs may well be biased against journals based in non-English speaking nations and this should be considered in any practical applications. More generally, the concept of research quality is not universal and may differ substantially between the Global North and Global South (e.g., Barrere, 2020). The JQF version used here employs the UK REF2021 definition of research quality and alternative versions that emphasised Global South priorities would produce different values that might be preferable for some research evaluations.

In the current climate, a counterintuitive practical advantage of JQFs is that they are clearly *not* research evaluations because they are based on titles and abstracts alone, ignoring article full texts. They might therefore be harder than average citation rates (e.g., JIFs) to mistake for direct evidence of research value. In theory, this might help inexperienced evaluators to use them with an appropriate level of caution and mitigate *some* of the problems of journal-based indicators that DORA has drawn attention to. Of course, this advantage will disappear if future LLMs are better able to process full texts and copyright allows it.